\documentclass[9pt,twocolumn,twoside]{pnas-new}

\templatetype{pnasresearcharticle} 

\graphicspath{{..}}

\begin{document}

\title{A game of life with dormancy}

\author[a]{Daniel Henrik Nevermann}
\author[a]{Claudius Gros}
\author[b]{Jay T.~Lennon}

\affil[a]{Institute for Theoretical Physics, Goethe University Frankfurt, Germany}
\affil[b]{Department of Biology, Indiana University, Bloomington, IN 47405, USA}

\leadauthor{Lead author last name}

\significancestatement{Authors must submit a 120-word maximum statement about the significance of their research paper written at a level understandable to an undergraduate educated scientist outside their field of speciality. The primary goal of the significance statement is to explain the relevance of the work in broad context to a broad readership. The significance statement appears in the paper itself and is required for all research papers.}

\authorcontributions{Please provide details of author contributions here.}
\authordeclaration{Please declare any competing interests here.}
\equalauthors{\textsuperscript{1}A.O.(Author One) contributed equally to this work with A.T. (Author Two) (remove if not applicable).}
\correspondingauthor{\textsuperscript{2}To whom correspondence should be addressed. E-mail: author.two\@email.com}

\keywords{Keyword 1 $|$ Keyword 2 $|$ Keyword 3 $|$ ...}

\begin{abstract}
The factors contributing to the persistence and stability of life are fundamental for understanding complex living systems. Organisms are commonly challenged by harsh and fluctuating environments that are suboptimal for growth and reproduction, which can lead to extinction. Species often contend with unfavorable and noisy conditions by entering a reversible state of reduced metabolic activity, a phenomenon known as dormancy. Here, we develop Spore Life, a model to investigate the effects of dormancy on population dynamics. It is based on Conway’s Game of Life, a deterministic cellular automaton where simple rules govern the metabolic state of an individual based on the metabolic state of its neighbors. For individuals that would otherwise die, Spore Life provides a refuge in the form of an inactive state. These dormant individuals (spores) can resuscitate when local conditions improve. The model includes a parameter $\alpha\in[0,1]$ that controls the survival probability of spores, interpolating between Game of Life ($\alpha=0$) and Spore Life ($\alpha=1$), while capturing stochastic dynamics in the intermediate regime ($0<\alpha<1$ ). In addition to identifying the emergence of unique periodic configurations, we find that spore survival increases the average number of active individuals and buffers populations from extinction. Contrary to expectations, the stabilization of the population is not the result of a large and long-lived seed bank. Instead, the demographic patterns in Spore Life only require a small number of resuscitation events. Our approach, yields novel insight into what is minimally required for the emergence of complex behaviors associated with dormancy and the seed banks that they generate.
\end{abstract}

\dates{This manuscript was compiled on \today}
\doi{\url{www.pnas.org/cgi/doi/10.1073/pnas.XXXXXXXXXX}}

\maketitle
\thispagestyle{firststyle}
\ifthenelse{\boolean{shortarticle}}{\ifthenelse{\boolean{singlecolumn}}{\abscontentformatted}{\abscontent}}{}

\firstpage[6]{4}

The text height is \the\textheight.
\dropcap{I}n nature, organisms are often challenged by conditions that are suboptimal for growth and reproduction. For life to persist, organisms must contend with a range of external forces. Scarcity of resources, fluctuating abiotic variables, and the patchy distribution of suitable habitat are just a few of the many exogenous factors that can reduce organismal fitness. In addition, individual performance is affected by endogenous factors arising from demographic rates and species interactions. Together, these factors pose a risk for local extinction. Populations can escape this fate through behavioral modifications, phenotypic plasticity, migration within a landscape, and evolutionary adaptation \cite{Parmesan_2006, Burton_etal_2022}. 

One process that is important in promoting the persistence of populations is dormancy, which occurs when an individual enters a reversible state of reduced metabolic activity \cite{Jones_Lennon_2010}. Dormant individuals enjoy protection against unfavorable conditions and have the capacity to resume growth when conditions improve. The accumulation of inactive individuals results in a "seed bank", which can buffer population dynamics leading to increased geometric mean fitness and a reduced probability of stochastic extinction in variable environments \cite{Menges_2000, Lennon_etal_2021}. 

Dormancy is achieved among species in different ways. In some cases, dormancy requires hundreds of interacting genes that are integrated into a tightly regulated developmental program \cite{McKenney_etal_2013}. Organisms adopting such a strategy often rely on the interpretation of environmental cues to transition between metabolic states in a responsive manner \cite{Kikuchi_etal_2022}. In other cases, dormancy is achieved through simpler means. For example, organisms can stochastically fall into an inactive state or randomly awaken independent of environmental conditions consistent with a bet-hedging strategy \cite{Kussell_Leibler_2005, Wright_Vetsigian_2019}. These fundamental differences in dormancy affect the size and longevity of a seed bank. While individuals belonging to some species only remain dormant for a short period of time \cite{Wilsterman_etal_2021}, individuals belonging to other species can persist indefinitely \cite{Brenner_etal_2013}. Such differences in shallow vs.\ deep dormancy should have implications for population stability and the persistence of life. 

Dormancy has independently arisen many times across the tree of life \cite{Lennon_etal_2021}. In this way, it is an example of convergent evolution \cite{Willis_etal_2014}, suggesting that dormancy may be a solution to one of life’s major problems, i.e., persisting in noisy and unpredictable environments. Despite its prevalence among diverse lineages and ecosystems, there is no standard way for modeling dormancy. While dynamical approaches have been developed \cite{Cohen_1966, tenBrink_etal_2020, Magalie_etal_2023} they typically do not capture the fact that dormancy is an individual process that can lead to complex emergent behaviors and patterns. Furthermore, modeling efforts have not identified the minimal requirements for achieving the benefits of dormancy, which could shed light on the origins and evolution of an important life-history strategy. 

One way to model living systems is through a bottom-up approach where rules are encoded into individual agents that are then observed as the consequences unfold over time. Perhaps the best examples come from cellular automata where self-replicating behaviors are governed by the agent’s state (e.g., $alive$ vs.\ $dead$) as well as those of its neighbors on a two-dimensional lattice with a random set of initial conditions. Despite their simplicity at the agent level, cellular automata, most notably Conway’s Game of Life \cite{Gardner_1970}, can give rise to periodic patterns \cite{Gotts_2000}, chaotic dynamics \cite{Nowak_May_1992}, self-organized criticality \cite{Bak_etal_1989}, and other life-like complexities \cite{Beer_2015}. While dispersal in space has been explored in cellular automata \cite{Yacoubi_Jai_2002,Hwang_etal_2009}, few studies have investigated how dispersal in time (i.e., dormancy) affects the dynamics of such systems \cite{Javid_teBoekhorst_2006, Locey_etal_2017}. 

Here, we develop a cellular automaton called Spore Life, which includes minimal modifications to the original Game of Life. We quantify the distribution of metabolic states along with the probability of extinction with updated rules that capture features of dormancy. In particular, we explored shallow vs.\ deep dormancy by manipulating a spore survivorship parameter, which also introduces stochasticity into the model. By characterizing the lifetime of individuals in active ($A$) and inactive ($I$) states, along with spore effects on demographic rates (i.e., births and deaths), we are able to recreate and better understand what is minimally required for dormancy and its contribution to population persistence. 

\section*{Methods}

\begin{figure*}[t]
    \centering
    \includegraphics[width=\textwidth]{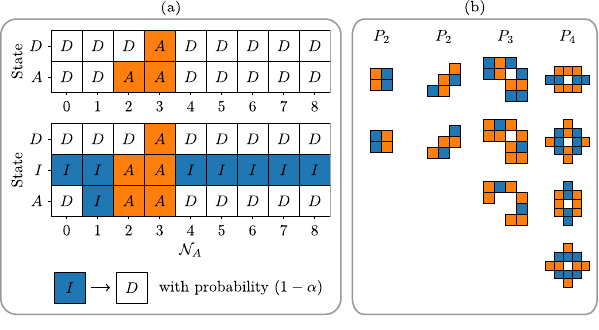}
    \caption{\textbf{Transition rules and unique configurations}
    In \textbf{(a)} we present rule tables for the Game of Life (top) and Spore Life (middle). Individuals can be in a dead ($D$), inactive ($I$), or active ($A$) state. Transitions among these states are governed by the number of active neighbors ${\mathcal{N}}_A\in \{1, 2,\dots,8\}$. Dormant individuals (i.e., spores) in an inactive state $I$ die at every time step with probability $1-\alpha$ (bottom). In \textbf{(b)} a selection of repeating configurations $P_n$ with period $n$ that arise in the deterministic limit ($\alpha=1$) of Spore Life. Active individuals are shown in orange and inactive individuals in blue.}
    \label{fig:rule-tables}
\end{figure*}

\paragraph{Game of Life}
We developed a cellular automaton that includes dormancy. The model is based on Conway’s Game of Life (GoL), which takes place on a two-dimensional lattice of sites $S$. In the original model, each site can be in dead state ($D$) or occupied by an alive individual in a metabolically active state ($A$) such that $S\in\{ \mathrm{active}, \mathrm{dead} \} \equiv \{A,D\}$. Initial conditions ($t = 0$) are created by seeding the $\ell\times \ell$ grid with probability $q$ of each site being occupied by an active individual. In subsequent time steps, the grid is updated in a density-dependent manner. Specifically, the following rules are applied to the occupant of a focal site based on the number of active individuals ($\mathcal{N}_A$) in neighboring sites (Fig.~\ref{fig:rule-tables}):  

\begin{enumerate}
    \item Active individual with $<$2 active neighbors dies ($A \rightarrow D$)
    \item Active individual with 2-3 active neighbors persists ($A \rightarrow A$)
    \item Active individual with $>$4 active neighbors dies ($A \rightarrow D$)
    \item Dead individual with 3 active neighbors is reborn ($D \rightarrow A$)
\end{enumerate}

\paragraph{Spore Life}
We created a model called Spore Life by introducing dormancy into the cellular automaton. This required the addition of an inactive state ($I$), allowing individuals to be in one of three metabolic states: $S \in \{\mathrm{active}, \mathrm{inactive}, \mathrm{dead}\} \equiv \{A, I, D\}$. As in the original Game of Life, initial conditions are created by seeding the $\ell \times \ell$ grid with a probability $q$ of each site being occupied by an active individual. In subsequent time steps, the grid is updated according to a slightly modified set of rules (Fig.~\ref{fig:rule-tables}):

\begin{enumerate}
    \item Active individual with 0 active neighbors dies ($A \rightarrow D$)
    \item Active individual with 1 active neighbor becomes a dormant spore ($A \rightarrow I$)
    \item Active individual with 2-3 active neighbors lives ($A \rightarrow A$)
    \item Active individual with $>$4 active neighbors dies ($A \rightarrow D$)
    \item Inactive individual (i.e., spore) with 2-3 active neighbors resuscitates ($I \rightarrow A$)
    \item Dead individual with 3 active neighbors is reborn ($D \rightarrow A$)
\end{enumerate}

\paragraph{Spore survivorship}
To explore how variation in spore survivorship influences population dynamics, we introduced a parameter $\alpha$, which determines the fate of an inactive individual (Fig.~\ref{fig:rule-tables} \text{a)}. We specify that an inactive individual dies with probability $(1-\alpha)$. As such, when $\alpha = 1$, an inactive individual has the potential to remain as a dormant spore for the remainder of the simulation. At the opposite limit, when $\alpha = 0$, all inactive individuals deterministically transition into dead individuals before the next time step. Under these conditions, the middle row of the Spore Life rule table is identical to the top row of the original Game of Life rule table (Fig.~\ref{fig:rule-tables} \text{a)}. When $\alpha \in (0, 1)$, transitions are stochastic such that some inactive individuals die while others resuscitate, reflecting variation in the degree to which organisms can persist in a shallow vs.\ deep state of dormancy. 

\paragraph{Dormancy and population dynamics}
Once implemented, we used Spore Life to characterize the effects of dormancy on key population-level phenomena. In addition to identifying the emergence of unique periodic configurations, we described how dormancy influences the abundance and distribution of metabolic states ($A$, $I$, $D$) on different sized grids. We also measured extinction probabilities of the population. While controlling for starting densities across a broad range ($n$ = 1,000) of initial condition, we quantified the number of time steps that occurred prior to extinction ($T_\mathrm{ext}$), which we operationally defined based on the constancy of $N_{A}$ after 100 time steps. 

\paragraph{Dormancy and demographic processes}
To gain mechanistic insight into how dormancy affected population-level phenomenon, we characterized two important demographic processes. First, we quantified the lifetime of inactive individuals for different levels of $\alpha$. This involved calculating the number of time steps before an inactive spore either died or was resuscitated. We then compared these estimates to the lifetime of an active individual, which involved measuring the number of time steps before it either died or transitioned into an inactive state $I$. Second, we examined how $\alpha$ contributed to the population by comparing sources of births (i.e., $D$ $\to$ $A$ vs.\ $I$ $\to$ $A$) and sources of deaths ($A$ $\to$ $D$ vs.\ $A$ $\to$ $I$) as a function of $\alpha$. 

An interactive web page is available to explore the dynamics of Spore Life: \url{https://itp.uni-frankfurt.de/spore-life/}

\section*{Results}

\paragraph{Spore Life creates unique periodic configurations}
In the Game of Life ($\alpha=0$) repeating configurations such as blinkers and gliders exist \cite{Brown_etal_2023}. This also holds in the opposite limit ($\alpha=1$) when Spore Life becomes deterministic. We identified a number of configuration that uniquely arise from the dormancy rule set. While our investigation of configurations was neither systematic nor exhaustive, we documented several patterns that repeat with periods of 2, 3, 4, 14, and 62 time steps (Fig.~\ref{fig:rule-tables}~\text{b} and Fig.~\ref{fig:sm-patterns}).

\paragraph{Dormancy stabilizes population dynamics}
When starting from a random distribution of metabolic states, we find that Spore Life is much more stable than the Game of Life (Fig.~\ref{fig:time-series}). Without dormancy ($\alpha = 0$), the number of active individuals ($N_A$) rapidly drops off leading to extinction in less than 100 time steps for a $\ell\times \ell =30 \times 30$ grid. Here, extinction occurs when there are no active or dormant individuals on the lattice, or when the population is comprised of isolated static or periodic configurations with densities of $\sim 3.3\%$ active individuals. In contrast, when dormant spores were generated due to local underpopulation, as illustrated in Fig.~\ref{fig:rule-tables} \text{a}, populations persist over prolonged time scales. In the deterministic limit, $\alpha = 1$, populations persisted for at least 1,000 time steps with much higher average population sizes. A simulation comparing the dynamics of Spore Life and the Game of Life can be viewed \href{https://https://itp.uni-frankfurt.de/spore-life/resources/spore-life.gif}{here}. In between, for $\alpha \in (0, 1)$, there was a reduction in the average $N_A$ relative to $\alpha=1$, suggesting an increased risk of local extinction (Fig.~\ref{fig:sm-alive-cell-distr})


\begin{figure*}[t]
    \centering
    \includegraphics[]{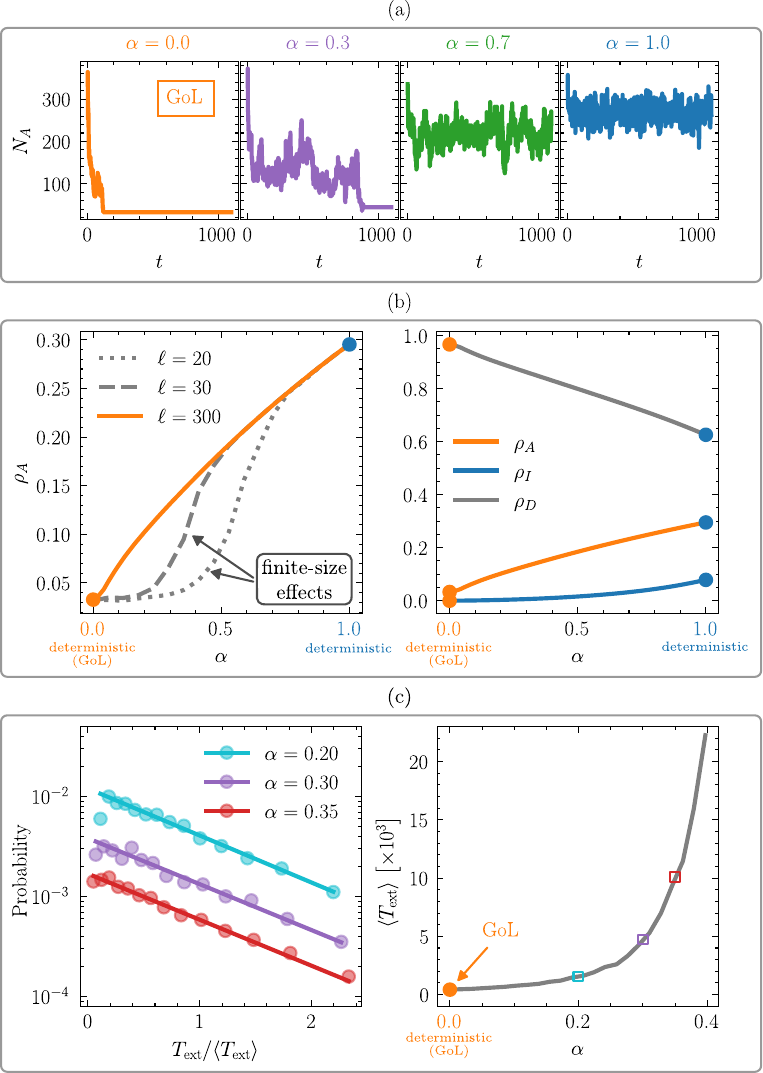}
    \caption{\textbf{Extinction times increase with dormancy} In Spore Life, we consider a population to be extinct if there are no active or dormant individuals, or if the pattern of active individuals consists only of isolated periodic or static configurations. From this, we define extinction time $T_\mathrm{ext}$ as the number of time steps starting from the random initial distribution of active individuals ($t = 0$) until an extinct state is reached. For a $\ell\times\ell=30\times30$ grid, we averaged over $10^3$ initial conditions, finding that the distribution of extinction times $T_{\rm ext}$ is exponentially distributed \textbf{(a)}. On average, extinction times rise sharply with increasing spore survivorship $\alpha$ \textbf{(b)}.}
    \label{fig:extinction-times}
\end{figure*}

\paragraph{Diminished effect of dormancy in small populations}
We identify a finite-size effect of dormancy on population dynamics. When spore survivorship ($\alpha$) is low, the stabilizing effect of dormancy on the average density of active individuals ($\rho_A = \langle N_A \rangle/\ell^2$) is weaker on smaller grids (e.g.\ $\ell = 20$ and $\ell = 30$) (Fig.~\ref{fig:finite-size-effects} \text{a}). Under such conditions, populations ($\langle N_{A}\rangle$) are smaller and experience larger fluctuations (Fig.~\ref{fig:sm-alive-cell-distr}), increasing the probability of extinction because locally static or periodic configurations remain isolated for all time. On a larger grid (e.g.\ $\ell = 300$), however, the average density of active individuals increases near linearly with $\alpha$ (Fig.~\ref{fig:finite-size-effects} \text{b)}. Under these conditions, where $\alpha \approx 1$, the population is well-mixed and individuals successively influence each other across all distances via the neighborhood-based update rules (Fig.~\ref{fig:rule-tables} \text{a)}.

\paragraph{Dormancy alters the distribution of metabolic states}
Spore survivorship alters the distribution of metabolic states in the population. Over the full range of $\alpha$, dormancy leads to an increase of $\rho_A$ from $\approx3\%$ for $\alpha=0$, to $\approx 30\%$ at $\alpha=1$ (Fig.~\ref{fig:finite-size-effects} \text{b}). At the same time, the average density $\rho_D$ of dead individuals decreases by $\approx35\%$. In contrast, the average density of inactive individuals $\rho_I$, which can be viewed as the "seed bank", remains low ($\rho_I<0.1$), even in Spore Life where $\alpha = 1$ (Fig.~\ref{fig:finite-size-effects} \text{b}). 


\paragraph{Dormancy reduces extinction probability}
Motivated by the temporal dynamics presented in Fig.~\ref{fig:time-series}, we sought to characterize the relationship between dormancy and the extinction probability. Considering a $30 \times 30$ grid, we found that extinction times ($T_\mathrm{ext}$) for a total of 1,000 simulations were distributed according to an exponential law $\sim c\,\mathrm{e}^{-T_\mathrm{ext}/T}$ (Fig.~\ref{fig:extinction-times} \text{a}). We find $c\; [\times 10^2]=1.2 / 0.38 / 0.17$ and $T = 0.92 / 0.95 / 0.95$ for $\alpha = 0.2 / 0.3 / 0.35$, respectively. While the decay constants ($T$) were essentially unaffected by spore survivorship ($\alpha$), dormancy had a strong effect on the scaling factors ($c$), which is evident from the different intercepts in Fig.~\ref{fig:extinction-times} \text{a}. The average time to extinction $\langle{T_{\mathrm{ext}}}\rangle$ increases exponentially with increasing spore survivorship. In the absence of dormancy ($\alpha = 0$), $\langle{T_{\mathrm{ext}}}\rangle$ was $\leq 500$ time steps, but even with moderate levels of spore survivorship ($\alpha = 0.4$), $\langle{T_{\mathrm{ext}}}\rangle$ already increased to $\geq 25,000$ time steps (Fig.~\ref{fig:extinction-times} \text{b}). Numerically, extinction times ($T_\mathrm{ext}$) increase with grid size. 


\paragraph{Spore survivorship minimally extends spore lifetimes}
To better understand how spore survivorship ($\alpha$) influences population dynamics in Spore Life, we characterized the distribution of spore lifetimes. Once created, an inactive individual's lifetime $T_I$ is equal to the number of time steps until a mortality event (when $\alpha < 1$) or a resuscitation event occurs. The probability that an isolated spore remains present after $t$ steps is
\begin{equation}
    \alpha^t = \mathrm{e}^{t\log(\alpha)}
    =\mathrm{e}^{-t/T_\alpha},
    \qquad  T_\alpha = \frac{-1}{\log(\alpha)}\,,
    \label{T_alpha}
\end{equation}
indicating that the lifetime of an inactive individual is exponentially distributed. 

\begin{figure*}[t]
    \centering
    \includegraphics[width=0.9\textwidth]{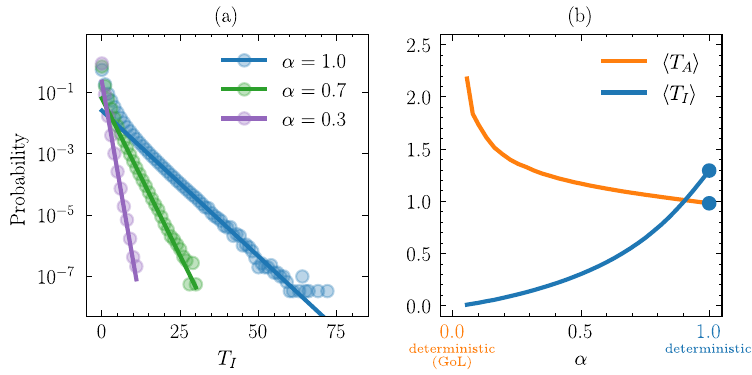}
    \caption{\textbf{Distribution of spore lifetimes}
    \textbf{(a)} The time steps ($T$) that an inactive individual remains in a dormant state ($D$) before dying or resuscitating into an active state ($A)$ decays exponentially. \textbf{(b)} The average life times of inactive individuals $\langle T_{I}\rangle$ and active individuals $\langle T_{A}\rangle$ as a function of $\alpha$.}
    \label{fig:spore-lifetime-distr}
\end{figure*}

Owing to a sharp initial drop off (Fig.~\ref{fig:spore-lifetime-distr} \text{a}), mean lifetimes are substantially shorter than the respective timescales $T$ obtained through fits (Fig.~\ref{fig:spore-lifetime-distr} \text{b}). Regardless of $\alpha$, the average lifetime of a spore ($T_{I}$) is $< 1.5$ time steps. Meanwhile, the average lifetime of an active individual ($T_{A}$) monotonically decreases with increasing $\alpha$ to about one time step for $\alpha\to1$ (Fig.~\ref{fig:spore-lifetime-distr} \text{b}). Although we did not numerically explore the exact behavior, it is worth noting that the lifetime of active individuals rapidly increases as $\alpha$ approaches 0. See Supporting Information for a discussion regarding the small values of $\alpha$.

From the exponential distribution, we conclude that the natural decay of spores $\sim(1-\alpha)$ is the dominant cause for the loss of inactive individuals when $\alpha$ is not too close to one. By examining the tails of (Fig.~\ref{fig:spore-lifetime-distr} \text{a}), which can be approximated by an exponential $\sim\exp(-t/T)$, we find $T = 0.77 / 2.14 / 4.56$ for $\alpha = 0.3 / 0.7 / 1.0$, whereas (\ref{T_alpha}) would predict $T_\alpha = 0.82 / 2.80 / \infty$, which is in turn reasonably close for $\alpha=0.3$ and $\alpha=0.7$. On the other hand, inactive individuals can only change to active individuals when $\alpha=1$, a process for which the typical timescale is observed to be $T=4.56$. 

\begin{figure*}[t]
    \centering
    \includegraphics[width=0.9\textwidth]{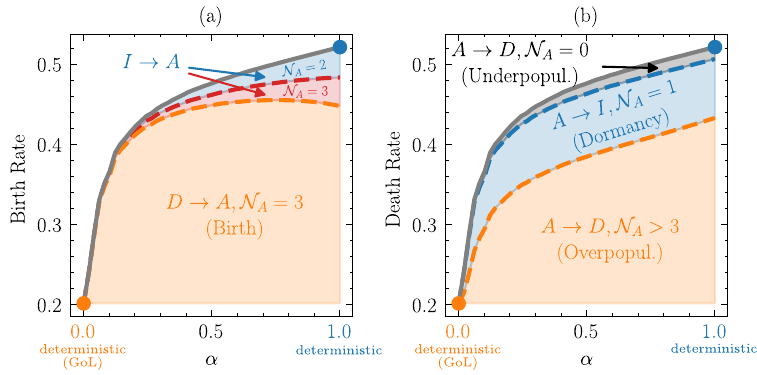}
    \caption{\textbf{Influence of dormancy on demographic processes} The gray lines correspond to \textbf{(a)} the birth rate and \textbf{(b)} the death rate of active individuals ($A$) as a function of spore survivorship ($\alpha$). The birth rate is defined as the number of transitions to the active state ($D \to A$ or $I \to A$) per time step normalized by the average number of active individuals $\langle N_A \rangle$, that is $\mathrm{Birth}\:\mathrm{Rate} = (\# \text{transitions to active}) / \langle N_A\rangle$. The death rate is defined as the number of transitions out of the active state ($A \to D$ or $A \to I$) per time step normalized by $\langle N_A\rangle$. For the birth rate, the primary source of new individuals comes from $D \to A$ transitions. Resuscitation events ($I \to A$) from individuals surrounded by either 2 or 3 active neighbors contribute equally, but rather minimally to the overall birth rate, especially at lower $\alpha$. For the death rate, the most dominant cause of death is overcrowding ($A \to D$ when $\geq 3$ active neighbors), while the contribution of underpopulation ($A \to D$ for 0 active neighbors) is relatively small. Dormancy ($A \to I$) plays an increasingly important role with increasing spore survivorship ($\alpha$).}
    \label{fig:birth-death-rate}
\end{figure*}

\paragraph{Spore survival alters demographic processes}
Spore survivorship ($\alpha$) substantially affects the birth rates and death rates of active individuals (Fig.~\ref{fig:birth-death-rate}). For small $\alpha$, $\approx20\%$ of the population dies and is reborn at every time step. As $\alpha\to1$, $\geq 50\%$ of the population dies and is reborn at every step. This is due to a higher likelihood of mortality from overcrowding (Fig.~\ref{fig:birth-death-rate} \textrm{b}) with increasing population density (Fig.~\ref{fig:finite-size-effects}). Somewhat counterintuitively, the probability of an active individual transitioning into the inactive state ($I$) increases substantially over the same range of {$\alpha$}. 

Regarding birth processes, at low- to intermediate-levels of $\alpha$, the vast majority of births ($\geq 90\%$) occur when active individuals are produced in an empty (dead) site that was surrounded by 3 active neighbors. As $\alpha \to 1$, $\approx13\%$ of new births are associated with the transition from an inactive state ($I$) to an active ($A$) state. As illustrated in Fig.~\ref{fig:rule-tables} \text{a}, these resuscitations happen when a spore is surrounded by either 2 or 3 active neighbors ($\mathcal{N}_A \in \{2, 3\}$). Our analysis suggests that the contributions of these two resuscitation pathways to the population birth rate are roughly equivalent (Fig.~\ref{fig:birth-death-rate} \text{a}). Regardless of whether an individual is in a dead $D$ or inactive $I$ state, when $\mathcal{N}_A = 3$, an active individual will be born. This means that any birth-mediated effect of dormancy can be attributed to the metabolic transitions that occur when $\mathcal{N}_A = 2$, which never accounts for more than $\approx8\%$ of the birth rate (Fig.~\ref{fig:birth-death-rate} \text{a}). Instead, at least $88\%$ of all new births are associated with $D \to A$ transitions, even when $\alpha=1$. 
 
In a similar way, we analyzed how different metabolic transitions contributed to the death rate of a population as a function of spore survival ($\alpha$). The largest source of mortality ($\approx80\%$) is due to overpopulation, which occurs when an individual is surrounded by $\geq 4$ active neighbors (Fig.~\ref{fig:birth-death-rate} \text{b}). In contrast, underpopulation is the smallest source of mortality ($\approx2\%$), which occurs when an active individual is surrounded by 0 active neighbors. Last, $\geq 18\%$ of the death rate was due to the loss of active individuals that transitioned into an inactive state $I$ when they were surrounded by 1 active individual.

\section*{Discussion}

Minor modifications to a simple cellular automaton were capable of capturing critical population-level phenomena that are commonly associated with dormancy. Using Spore Life, we demonstrate that dormancy confers stability, defined as population persistence over time (Figs.~\ref{fig:time-series}, ~\ref{fig:sm-alive-cell-distr}). To address the effects of shallow vs.\ deep dormancy, we included a spore survivorship parameter ($\alpha$) that also introduces stochasticity into the model. As expected, extinction times increase with spore survivorship (Fig.~\ref{fig:extinction-times}). Unexpectedly, the benefits of dormancy do not require the accumulation of inactive individuals into a large or long-lived seed bank. Instead, dormancy promotes population persistence through slight increases in spore lifetimes (Fig.~\ref{fig:spore-lifetime-distr}), modest contributions to the metabolic processes regulating demographic rates (Fig.~\ref{fig:birth-death-rate}), and likely feedbacks created by the emergence of unique periodic configurations (Fig.~\ref{fig:rule-tables} \text{b)}. Our findings reveal that small changes to a relatively simple cellular automaton can yield complex behaviors that are consistent with theoretical expectations of how dormancy stabilizes populations in fluctuating environments. In the following sections, we discuss these findings in the context of what is known about dormancy from other computational models and how our results may guide future research efforts, including empirical studies relating to resiliency of host and environmental health. 

\paragraph{Scale-dependence of dormancy}
Dormancy is governed by a range of distinct processes that may operate across spatial and temporal scales \cite{Jones_Lennon_2010}. Some drivers of dormancy occur seasonally and integrate over large spatial extents (e.g., photoperiod) \cite{Saunders_2020}. Dormancy can also be triggered by conditions that arise at the local scale, including density-dependent fluctuations in the abundance of individuals belonging to a cohort that experience intraspecific competition for a limiting resource \cite{Nilsson_etal_1994}. In Spore Life, dormancy dynamics are entirely controlled by the metabolic transitions of individuals and their nearest neighbors, and not because of external forces. These local interactions then play out over the entire landscape and determine the emergent dynamics of the population. 

One scale-dependent phenomenon observed in our study relates to the effect of dormancy in habitats (i.e., grids) of different sizes. In the Game of Life, which lacks dormancy ($\alpha = 0$), the proportion of active individuals ($\rho_A$) rapidly declines over time (Fig.~\ref{fig:time-series}). Consistent with an Allee effect \cite{Stephens_etal_1999}, a critical point is reached where local neighborhoods are so depleted of active individuals that new individuals cannot be born ($D \to A$), and the population effectively goes extinct. In Spore Life, where $\alpha>0$, the opposite is true, in particular when $\alpha$ is large ($\alpha\to1$). The proportion of active individuals is sustained at a much higher level (Fig.~\ref{fig:time-series}) owing to the dormancy refuge ($A \to I$) and resuscitation events ($I \to A$). In the two deterministic limits ($\alpha = 0/1$), the effects of dormancy are consistent across simulations independent of grid size. However, at intermediate levels of spore survivorship, the effects of dormancy are more variable. Specifically, on small grids with low- to intermediate-levels of $\alpha$, there is a higher probability for elementary configuration to become fixed, which leads to low $N_{A}$ and subsequent extinction. In contrast, populations persist longer with low- to intermediate-levels of $\alpha$ on larger grids. This finite-size effect may reflect specific features that are unique to our cellular automaton. Alternatively, the model may be capturing a more general phenomenon -- albeit one that is not well-documented -- where smaller populations, perhaps in more fragmented habitats, are less likely to be rescued by dormancy.

\paragraph{Shallow vs.\ deep dormancy}
Dormancy is a process that can lead to the formation of a seed bank, a subpopulation of metabolically inactive individuals. There are many seed-bank attributes that can affect the ecology and evolution of a population \cite{Lennon_etal_2021}. One important attribute is seed bank size. A large seed bank comprised of many inactive individuals provides more opportunities for resuscitating individuals to buffer a population against extinction. Another important attribute is seed bank turnover, which is influenced by the amount of time that dormant individuals can survive while in a metabolically inactive state. Longer-lived dormant individuals should provide more insurance to a population in a fluctuating environment, provided that they remain capable of resuscitation. At the same time, prolonged dormancy can be maladaptive because as it delays growth and reproduction \cite{Gremer_et_al_2012}. These general hypotheses led us to introduce the parameter $\alpha$ into our model to explore how spore survivorship contributes to the outcomes of dormancy. We found that the effects of dormancy on the population were strongly affected by $\alpha$, namely that the abundance of active individuals ($N_{A}$) increases with $\alpha$, with the probability of extinction decreasing with $\alpha$. 

Contrary to our expectations, spore survivorship ($\alpha$) did not have a strong effect on seed bank size (i.e., $\rho_I$), or on the lifetime of a spore ($T_{I}$). Instead, the effects of dormancy were achieved by individuals that spent only a short time in an inactive state ($I$). The average lifetime of a spore was quite low, on the order of only a single time step (or generation). Meanwhile, the contribution of resuscitation ($I \to A$) to the overall birth rate of active individuals was low ($< 12\%$) compared to births not directly associated with dormancy ($D \to A$). On the larger grids (e.g.\ $300 \times 300$), the density of inactive individuals ($\rho_I$) never comprised more than 8\% of the grid or 22\% of the viable (i.e., active + inactive) individuals. Seed bank size, defined as $N_I$:$N_A$, increased with $\alpha$, but was generally low (0.02 - 0.27). While turnover increased with $\alpha$, resuscitation with $\mathcal{N}_A = 2$ living neighbors never accounted for $> 11\%$ of all births. 

One explanation for population persistence in Spore Life (Fig.~\ref{fig:time-series}) is sustained chaotic transients. In complex systems, the introduction of noise can lead to superpersistent transient chaos \cite{Do_Lai_2005}. Specifically, lattice-wide noise could be generated in simulations with stochastic spore survivorship ($\alpha \in (0, 1)$). However, this explanation is not consistent with observed behavior in the deterministic limit of Spore Life ($\alpha = 1$) where we see clear stabilization of populations induced by dormancy.

Another possibility is that dormancy may be important for population persistence, even if inactive individuals are short-lived and seed banks are relatively small. Such ideas are not well-developed in theoretical treatments of dormancy, although biologists have noted "cryptic" forms of metabolic stasis in some groups of taxa \cite{Doropoulos_etal_2022, Polačik_Vrtílek_2023}. Thus, our findings suggest there may be relatively unexplored regimes where dormancy may be important for stabilizing populations, but this requires the simultaneous development of theory and empirical approaches. 

\paragraph{How "easy" is dormancy?}
Dormancy has evolved independently many times throughout the tree of life. It would be impossible, therefore, to create a single model representing all features and attributes of dormancy as a diverse life-history strategy. As such, our goal was to develop a compact model with the fewest number of rules and assumptions. We find that Spore Life is successful in recapitulating a range of canonical behaviors that are associated with dormancy, in particular with regard to the dampening of population dynamics in a fluctuating environment \cite{Caceres_1997}. 

That being said, there are other seed-bank attributes that could be incorporated into future versions of Spore Life. In nature, dormant and active individuals are typically not well-mixed \cite{Lennon_etal_2021}. For example, much of an actively growing plant is found aboveground while dormant seeds, after being dispersed from their natal sites, often reside belowground in the soil. This population structure could be represented in the metabolic-transition rules of Spore Life (Fig.~\ref{fig:rule-tables} \text{a)}. Also, there are ways that coarser-grained heterogeneity could be encoded onto a grid,  or even a 3D spatial lattice of a cellular automaton \cite{Bays_2010, Glade_etal_2017}, that would allow for tests of seed bank dynamics in a more complex universe. One might expect that the spatial decoupling of dormant and active individuals would reduce the reactivity of the system and give rise to larger and longer-lived seed banks. 

Furthermore, in the current version of Spore Life, there is only a single condition when dormancy acts as a refuge (Fig.~\ref{fig:rule-tables}), which occurs due to underpopulation when an active individual is surrounded by a single active individual. However, initiation of dormancy could also be triggered by crowding \cite{Habtewold_etal_2021}, for example, when an active individual is surrounded by $ \geq 4$ four active individuals (see Supporting Information). 

Last, While we included stochasticity in the context of spore survivability, there are opportunities to explore how mixtures of random and deterministic transitions between metabolic states affect population dynamics \cite{Blath_Andras_2020}. Such efforts would allow for exploration and synergy with other applications of dormancy-related research, including those focused on evolutionary behaviors such as adaptive dynamics \cite{Blath_etal_2023}, interacting particle systems in statistical physics \cite{Floreani_etal_2022}, and the treatment of diseases, such as chronic infections \cite{Rittershaus_etal_2013} and cancer \cite{Aguirre-Ghiso_2007}. Ultimately, the platform developed here has the potential to shed light on the origins and diversification of a widespread trait characterized by complex spatial and temporal dynamics, together with the associated metabolic feedback loop.

\section*{Supporting Information}\label{sec:suppl}
\renewcommand{\thefigure}{S\arabic{figure}}
\setcounter{figure}{0}

\begin{figure*}[t]
    \centering
    \includegraphics[scale=0.8]{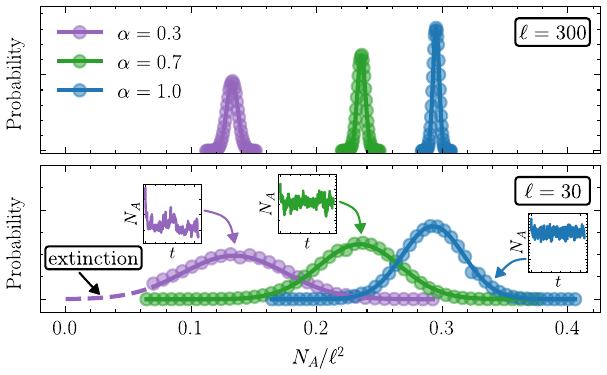}
    \caption{\textbf{Distribution of the population size} For different spore survival probabilities $\alpha$, we constructed the distribution of active population densities, $N_A/\ell^2$, where we considered $\ell = 300$ (top) and $\ell = 30$ (bottom). Shown are least-square Gaussian fits (lines) to the data (circles), where in the smaller grid ($\ell = 30$) we find the means $\mu = 0.13 / 0.23 / 0.29$ and standard deviations $\sigma = 0.041 / 0.032 / 0.024$ for $\alpha = 0.3 / 0.7 / 1.0$, respectively. In the larger grid ($\ell = 300$), means are unchanged and standard deviations are smaller by a factor 10. The insets show examples for the respective time series. For $\ell = 30$ and $\alpha = 0.3$ the distribution overlaps with unstable densities which cause fluctuation induced extinction (dashed line).}
    \label{fig:sm-alive-cell-distr}
\end{figure*}

\paragraph{Population fluctuations are Gaussian}
In Fig.~\ref{fig:time-series} and \ref{fig:finite-size-effects} we demonstrate how population dynamics change in response to spore survival probability ($\alpha$). Increasing $\alpha$ leads to larger average population sizes containing larger numbers of active individuals ($N_A$). This pattern is accompanied by decreased fluctuations in population size, which are Gaussian distributed (Fig.~\ref{fig:sm-alive-cell-distr}).

The means of the fitted Gaussians $\langle N_A \rangle$ increase near linearly with the spore survivability $\alpha$, as demonstrated in Fig.~\ref{fig:finite-size-effects}~b. We further find that the standard deviation decreases roughly linearly with $\alpha > 0$, where the treatment at small $\alpha\sim0$ is numerically demanding (see below). For smaller grids (e.g., $\ell = 30$) fluctuations are much broader than in larger grids (e.g., $\ell = 300$), where the observed standard deviations are scaled by a factor of 10. Importantly, for small $\alpha$ a high standard deviation may cause fluctuation induced extinction events, when the distribution overlaps with densities of actives cells that are unstable towards extinction (Fig.~\ref{fig:sm-alive-cell-distr}, dashed line).

\paragraph{Lifetimes of active and inactive cells close to GoL} In Fig.~\ref{fig:spore-lifetime-distr}~b we did present the average lifetimes of active and inactive cells, $\langle T_A\rangle$ and $\langle T_I\rangle$ respectively. Here we expand upon the behavior close to the transition to the deterministic GoL ($\alpha$ small). For inactive cells, it is clear that the average lifetime $\langle T_I \rangle = 0$ at  $\alpha = 0$ for $\alpha \to 0$, as spores die in the GoL limit (Fig.~\ref{fig:rule-tables}). The average lifetimes of active individuals, $\langle T_A\rangle$, are however not well-defined when $\alpha \sim 0$. In this parameter regime, the average extinction time, $\langle T_\mathrm{ext}\rangle$, is low (Fig.~\ref{fig:extinction-times}~b). According to our definition, an extinct system may contain active (and inactive) individuals in stable configurations or blinkers, thus active individuals in an extinct state may either stay active for eternity (or rather for the duration of the simulation), or periodically die after a single time step. Taking the average after population extinction therefore seems unreasonable. As a consequence, as extinction times are short for small $\alpha$, there is no window for which a stable transient state may be defined. It is hence not possible to measure, or to define, the average lifetime of active cells in the GoL limit.

\paragraph{Overpopulation induced dormancy}
In unfavorable environmental conditions an individual may enter a dormant state with reduced metabolic activity, which therefore increases the endurance of life. External factors challenging organisms can be of very different nature. So far, we took the neighborhood of a living cell as a predictor for the quality of the environment (Fig.~\ref{fig:rule-tables}), where we assumed that harsh conditions, and thus an impetus for dormancy, are present when the number of active neighbors was low. Nevertheless, external factors challenging the survival of individuals could arise also in the case of overpopulation, e.g.\ when resources or habitat become scarce as a result of high population densities. Overpopulation is hence an alternative route for dormancy to arise. As a robustness check we implemented a modified version of Spore Life compared to Fig.~\ref{fig:rule-tables}, where a further transition to the inactive state ($I$) is included for active individuals with $\mathcal{N}_A = 4$ living neighbors, viz for mild overpopulation. While we find that absolute figures change in this modified model, the aforementioned phenomenology remains valid.

\begin{figure*}[t]
    \centering
    \includegraphics[]{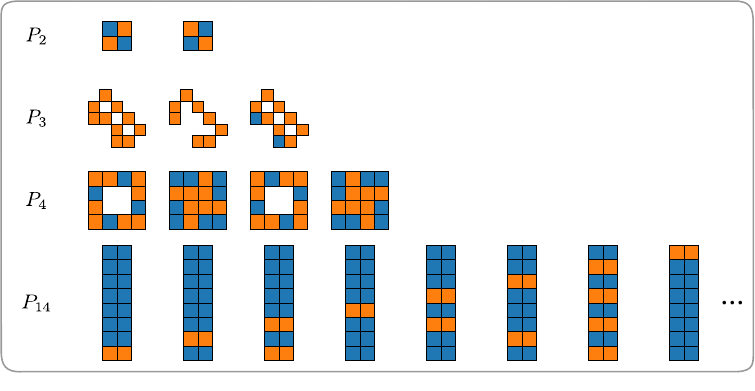}
    \caption{\textbf{Patterns} Additional to Fig.~\ref{fig:rule-tables}~b we present periodic patterns $P_n$ of periods $n = 2, 3, 4$ and $14$ found the deterministic Spore Life ($\alpha = 1$). See \url{https://itp.uni-frankfurt.de/spore-life/} to interactively explore more periodic patterns.}
    \label{fig:sm-patterns}
\end{figure*}

\acknow{We acknowledge support from the US National Science Foundation (DEB 1934554 and DBI-2022049 to JTL), the US Army Research Office Grant (W911NF2210014 to JTL), the National Aeronautics and Space Administration (80NSSC20K0618 to JTL), and the Alexander von Humboldt Foundation (to JTL). We also acknowledge Pat Wall for discussion during the early stages of this project, along with Bulcsú Sándor and Kevin Webster for helpful comments. Code for the model and analyses is available on GitHub: \url{https://github.com/ISCOTTYI/spore-life}}

\showacknow{} 

\bibsplit[2]

\bibliography{gol}

\end{document}